\documentclass[10pt,journal,twoside,compsoc]{IEEEtran}
\ifCLASSOPTIONcompsoc
   \usepackage[nocompress,noadjust]{cite}
\else
   \usepackage[noadjust,sort]{cite}
\fi
\ifCLASSINFOpdf
  \usepackage[pdftex]{graphicx}
\else
\fi
\usepackage[cmex10]{amsmath}
\usepackage{amsfonts}
\usepackage[all]{xy} 
\usepackage{pb-diagram,pb-xy}
\ifCLASSOPTIONcompsoc
\usepackage[hang,raggedright,tight,footnotesize,sf,SF]{subfigure}
\else
\usepackage[hang,raggedright,tight,footnotesize]{subfigure}
\fi
\usepackage{url}

\usepackage{multirow}
\usepackage{array}
\newcolumntype{V}{>{\centering\arraybackslash} m{.4\linewidth} }
\DeclareMathOperator{\System}{System}
\DeclareMathOperator{\Data}{Data}
\DeclareMathOperator{\Representation}{Representation}
\DeclareMathOperator{\Visualization}{Evocation}
\DeclareMathOperator{\Knowledge}{Knowledge}
\DeclareMathOperator{\Schema}{Schema}
\DeclareMathOperator{\Layout}{Layout}

\DeclareMathOperator{\Question}{Questions}
\DeclareMathOperator{\A}{A}
\DeclareMathOperator{\B}{B}
\DeclareMathOperator{\C}{C}

\DeclareMathOperator{\Obj}{O}
\DeclareMathOperator{\R}{R}
\DeclareMathOperator{\I}{I}
\DeclareMathOperator{\X}{X}

\begin{document}
%
\title{Understanding Visualization:\\A Formal Approach using Category Theory and Semiotics}
%
%
%
%

\author{Paul~Vickers,~\IEEEmembership{Member,~IET}
        Joe~Faith,
        and~Nick~Rossiter.
\IEEEcompsocitemizethanks{\IEEEcompsocthanksitem P. Vickers and N. Rossiter are with the School of Computing, Engineering, and Information Sciences, Northumbria University, Newcastle-upon-Tyne, NE2 1XE, United Kingdom.\protect\\
E-mail: paul.vickers@northumbria.ac.uk
\IEEEcompsocthanksitem Joe Faith is with Google Corp. Inc., Mountain View, CA 94043, USA.}
\thanks{Manuscript received ; revised .}}

%
%

\markboth{IEEE Transactions on Visualization and Computer Graphics,~Vol.~x, No.~x, month~year}%
{Vickers \MakeLowercase{\textit{et al.}}: Understanding Visualization}
%

\IEEEpubid{\makebox[\columnwidth]{\hfill 0000--0000/00/\$00.00~\copyright~2011 IEEE}%
\hspace{\columnsep}\makebox[\columnwidth]{Published by the IEEE Computer Society\hfill}}


\IEEEcompsoctitleabstractindextext{%
\begin{abstract}
This article combines the vocabulary of semiotics and category theory to provide a formal analysis of  
visualization. It shows how familiar processes of 
visualization fit the semiotic frameworks of both Saussure and Peirce, and extends these structures using the tools of category theory to provide a general framework for understanding visualization in practice, including: relationships between systems, data collected from those systems, renderings of those data in the form of representations, the reading of those representations to create visualizations, and the use of those visualizations to create knowledge and understanding of the system under inspection. The resulting framework is validated by demonstrating how familiar information visualization concepts (such as literalness, sensitivity, redundancy, ambiguity, generalizability, and chart junk) arise naturally from it and can be defined formally and precisely. This article generalizes previous work on the formal characterization of 
visualization by, \emph{inter alia}, Ziemkiewicz and Kosara and allows us to formally distinguish properties of the visualization process that previous work does not.
\end{abstract}

}

\maketitle

\IEEEdisplaynotcompsoctitleabstractindextext

 \ifCLASSOPTIONpeerreview
 \begin{center} \bfseries EDICS Category: 3-BBND \end{center}
 \fi
%
\IEEEpeerreviewmaketitle

\section{Introduction}
%
%

%
%
%
%
\IEEEPARstart{V}{isualization} is a catch-all term that embraces a wide range of activities concerned with representing, or making visible aspects or features of a given set of data or system, from the graphical analysis of scientific data, through the `infographics' used to  communicate in the popular media, to data art. It has recently grown in scale, popular currency, and theoretical discussion due to a combination of factors including the growth in the importance of data mining and processing in industry and science, and the availability of popular and powerful computer visualization tools, such as Processing (see \cite{Processing}). 

Visualization practice combines a range of skills and disciplines, including statistics, aesthetics, HCI, graphic design, and computer science. However, perhaps because of this diversity, there has been relatively little discussion of its theoretical basis. Purchase et al. \cite{Purchase:2008} remarked that 
visualization ``suffers from not being based on a clearly defined underlying theory'', and that ``formal foundations are at a nascent stage''. The danger of neglecting the theoretical foundations is that the discipline will fragment into isolated communities of practice that fail to learn from one another and replicate work unnecessarily. There have been a number of efforts to map visualization's theoretical foundations, the principal ones being Mackinlay \cite{Mackinlay:1986}, Card and Mackinlay \cite{Card:1997}, Chi \cite{Chi:2000}, Tory and M{\"o}ller \cite{Tory:2004},  van Wijk \cite{vanWijk:2006} and Ziemkiewicz and Kosara \cite{Ziemkiewicz:2009} (see also \cite{Kosara}). We discuss the relationship between these treatments and ours in section \ref{sec:theory}  and  show how using a more powerful formal framework allows us to formally distinguish properties of the visualization process that previous work does not.  

This article contributes to this development of theoretical foundations by generalizing previous work and providing a higher-level framework for understanding the visualization process. It does this by employing two existing tools. The first is semiotics, the study of signs. As devised by Saussure and Peirce, semiotics has developed into a powerful theoretical framework for understanding the relationships between signs, sign systems, the consumers of those signs, and the systems they represent. Information visualizations are signs \emph{par excellence}, and thus seem obvious candidates for semiotic analysis. In section \ref{sec:semiotics} we provide a brief introduction to the analysis frameworks used by semiotics and discuss how this applies to  
visualization. The result is to show how  
visualization can be understood using a series of relationships, or mappings, from one domain to another, summarized in a semiotic triad.

The second tool is category theory, the mathematical study of systems of structures and their mappings. This is introduced in section \ref{sec:CT}, and applied to  
visualization in section \ref{sec:CTVis}. 
The most powerful concept in category theory is the notion of commutativity, which forces one to try to extend and construct structures in such a way as to reach algebraic closure by considering the consequences and implications of a structural description of a system. The end result, or closure, is the general description of  
visualization given in section \ref{sec:closure} culminating in Fig. \ref{fig:Process} and Table \ref{tab:morphSummary}. By applying the criteria of commutativity and closure to Peirce's semiotic triad in sections \ref{sec:CTVis} and \ref{sec:application} we use our framework in a natural way to cover (or uncover) some aspects which are already familiar to practitioners of information visualization, and some other aspects which are less obvious. Specifically, the theory lets us differentiate between chart junk that is part of the schema, that which is arbitrary, and redundancy in the layout. We also show how the derived property of `intensionality' can be used to discriminate between true visualizations and data-driven representations.  

\section{Semiotics and Visualization}
\label{sec:semiotics}
Semiotics is the study of the creation and interpretation of signs. Signs are words, images, sounds, smells, objects, etc. that have no intrinsic meaning and which become signs when we attribute meaning to them \cite{Chandler:2007}. Signs stand for or represent something beyond themselves. Modern semiotics is based upon the work of two principal thinkers, the Swiss linguist Ferdinand de Saussure and the American philosopher and logician Charles Sanders Peirce. In Saussure's linguistic system a sign is a link between the \emph{signified} (a concept) and the \emph{signifier} (a sound pattern) both of which are psychological constructs having non-material form rather than material substance \cite{Chandler:2007}. For example, /tree/ is a signifier for the concept of the thing we know as a tree. The sign thus formed is a link between the sound pattern and the concept. Modern applications admit material form for the signifier (e.g., road signs and printed words). Saussure's scheme explicitly excludes reference to objects existing in the real world; the signified is not directly associated with an object but with a mental concept.

\begin{figure}[htbp]
  \centering
      \includegraphics{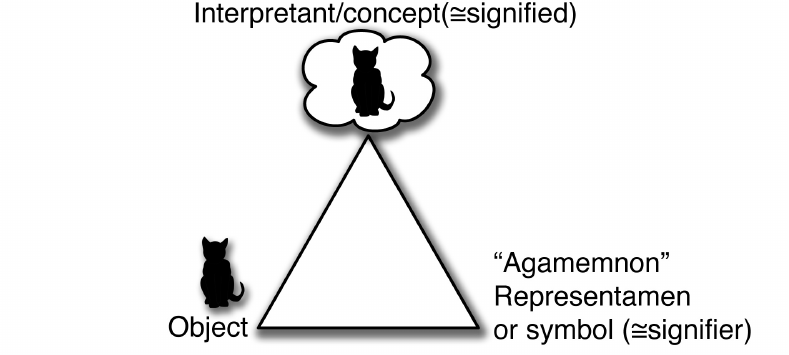}
    \caption{Semiotic `meaning triangle' showing a Peircean semiotic triad (notation after Sowa \cite{Sowa:2000}). Approximations to Saussure's  terms are given in parentheses: A real-world cat is the object. It is represented by the symbol ``Agamemnon'' which evokes the concept of Agamemnon the cat in our mind.}
    \label{fig:triangle}
\end{figure}

Peirce's semiotics is based upon a triadic relationship that comprises \cite{Vickers:2011}: 
\begin{itemize}
	\item The \emph{object}: the thing to be represented (note, this need not take a material form); 
	\item The \emph{representamen}: the form that the sign takes (word, sound, image, etc.,); the representamen represents the object; 
	\item The \emph{interpretant}: the sense we make of the sign. 
\end{itemize}
Peirce thus admits the referent that Saussure brackets. Fig.~\ref{fig:triangle} (adapted from Vickers \cite{Vickers:2011}) shows a Peircean triad drawn as a `meaning triangle'.\footnote{In 1923 Ogden and Richards \cite{Ogden:1923} gave this visual interpretation of Peirce's triad since when it has become the conventional representation (see also Sowa \cite{Sowa:2000}).} It should be noted that the Saussurean signifier and signified correspond only approximately to Peirce's representamen and interpretant; unlike Saussure's signified, Peirce's interpretant itself becomes a sign vehicle in the mind of the interpreter. The triangle shows the  sign formed by the name Agamemnon which signifies an individual cat of that name.

\begin{figure}[htbp]
  \centering
    \includegraphics{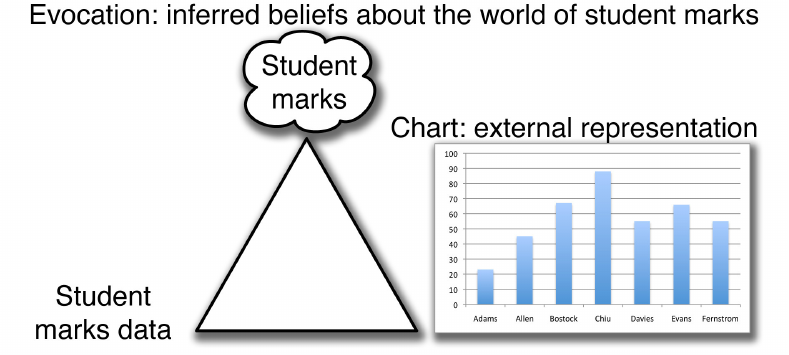}
      \caption{A chart is a representation of the data. The interpretant is the concept formed in our mind when we read the chart representation.}
      \label{fig:chart}
\end{figure}

To see how this relates to visualization consider Fig.~\ref{fig:chart} (adapted from Vickers \cite{Vickers:2011}) which shows a semiotic relationship between a set of student marks and an external representation. Using a spreadsheet program we take a data set collected from the real world system of a cohort of students studying a course. These data are then presented to the user as a chart. Note, the chart is not the data set but a representation of it.  So, we have a Peircean sign in which the data set is the referent object, the chart serves as its representamen, and the interpretant is the sense we make of the data by studying the chart.

It is important to note that, contrary to Saussure's original structuralist view, sign systems exist within a social and cultural context which, the post-struct\-ur\-alists would argue, needs to be accounted for. Peirce's semiotics, through the notion of a ground, admits such a context. It is important because visualization requires the producer (the addresser, in semiotic terminology)
and the consumer (the addressee) 
to share contextual knowledge for successful meaning making to take place (see also van Wijk \cite{vanWijk:2006}). In Peirce's semiotics meaning is mediated such that the ``meaning of a sign is not contained within it, but arises in its interpretation'' \cite{Chandler:2007}. Hjelmslev \cite{Hjelmslev:1961} recognized that no sign can properly be interpreted without first contextualizing it so that in addition to a sign's denotative (literal) meaning its context also lends it connotative meaning. For instance, in the example of our student system the ground would include knowledge about what constitutes a pass mark  and where the grade boundaries lie (see also Hullman and Diakopoulos \cite{Hullman:2011}). 

Note that this semiotic framework excludes some examples of what is popularly regarded as information visualization such as Radiohead's ``House of Cards'' video \cite{Radiohead:2007}. The video was shot without any cameras, being derived solely from data obtained from 3D images produced by Geometric Informatics for close proximity objects and Velodyne Lidar for landscapes. The data sets used to make the video can be downloaded from the project's web site which encourages us ``to create your own visualizations'' \cite{Radiohead:2007}. According to Lima \cite{Lima:2009}, the video, while unarguably data-driven, ought not be considered a visualization as it provides no insight into the data --- it is pure spectacle \cite{Barrass:2011}. As Card, Mackinlay, and Shneiderman put it: ``The purpose of visualization is insight, not pictures'' \cite{Card:1999}. 
Visualization, then, is a process that begins with the real world, or more narrowly, a system in the real world about which we are interested (e.g., a mechanical system or a cohort of students on a degree course). From the system we gather data which are then mapped via transformation rules to a representation (a graph or chart, an interactive 3D model, statistical box plots, etc.) and this representation is then `read' by the person wishing to gain insight into the system. Note that the data being visualized always have a context (a ground), a scenario in which they were intended for use. 

\begin{figure}[htbp]
  \centering
    \includegraphics{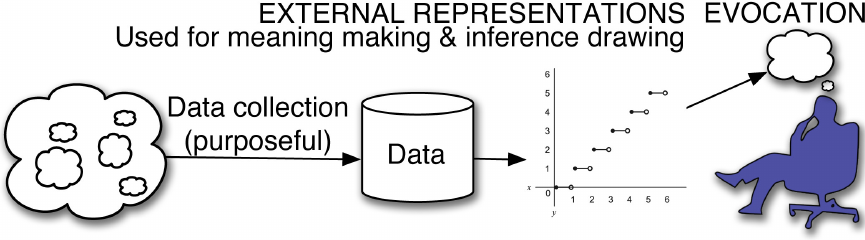}
        \caption{Understanding the real world through visualization: Data are purposefully collected from the real world and, via mappings, representations are produced. These are used in turn for meaning making and drawing inferences about the data. The visualization process encompasses cognition in the observer's mind.}
    \label{fig:SonVis}
\end{figure}

Reading the representation evokes concepts and ideas in the mind and inferences are drawn leading to understanding of the system and, as we shall see later, knowledge of the truth as it pertains to that system. This process is encapsulated in Fig.~\ref{fig:SonVis}.

\section{Category Theory}
\label{sec:CT}
In this section we introduce the main concepts of category theory \cite{MacLane:1971}, a branch of mathematics developed to analyze systems of structures, and mappings between those structures, in their most general form. This level of generality means that category theory is capable of demonstrating similarities between disparate fields of mathematical enquiry --- from set theory to theoretical computer science --- in such a way that allows insights from one to be translated to another. It can thus act as a Grand Unified Theory in mathematics.

In sections \ref{sec:MOT}--\ref{sec:morphisms} we introduce the basic elements and concepts of category theory, and in sections \ref{sec:VisProcCat}--\ref{sec:closure} we show how these elements and concepts apply to visualization processes. In section \ref{sec:application} we derive some results of analyzing information visualization in this way.

\subsection{Morphisms, Objects, and Triangles}
\label{sec:MOT}
Here we introduce the central concepts of the object and the morphism and
 how they may be arranged into diagrams, the fundamental one being the triangle. 
In section \ref{sec:CTVis} we will see that the elements of a visualization process (such as data sets and representations) are objects and the processes that map between them (such as measurement and rendering) are morphisms.

Category theory is built from just two classes of entity: objects and morphisms (we follow the convention that Objects are Capitalized and \emph{morphisms} are \emph{italicized}). Almost anything can be considered as an object: physical objects, abstract objects, or entire systems. Indeed, much of category theory's power comes from its recursive ability to treat ever more complicated systems as building blocks in the next level of abstraction. All that is required for an entity to qualify as an object is that it can be individuated, that is, we have some method for determining whether two objects are identical.\footnote{In category theory it is strictly not possible to show that things are the same or identical. The strongest statement possible is that two sets are naturally isomorphic (unique up to natural isomorphism), that is, indistinguishable.} This method is represented as a mapping from the object to itself, known as the identity morphism (see section \ref{sec:CIA}).

Morphisms are mappings between objects. They are represented diagrammatically by (and often called) arrows. For a mapping to qualify as a morphism there must be a unique target for each domain object, i.e., one object each at the base and the head of the arrow. Ontologically, a morphism may be understood as a generalization of a mathematical function, i.e., as an association between its source and target. Morphisms may represent physical, causal, or temporal processes, or purely formal relationships.

\begin{figure}[htbp]\centering
    \subfigure[The basic triangle]{
     $$\xymatrix{\A\ar[r]^f\ar[dr]_{h}&\B\ar[d]^{g}\\
	&\C
	}$$
   \label{fig:basicTriangle}} \quad\quad
      \subfigure[Triangle that commutes]{
       $$\xymatrix{\A\ar[r]^f\ar[dr]_{g\circ f}&\B\ar[d]^{g}\\
	&\C
	}$$
      \label{fig:commutingTriangle}}
    \caption{Category theory triangles. If the triangle in (b) commutes then $h=g\circ f$.}
    \label{fig:triangles}
\end{figure}
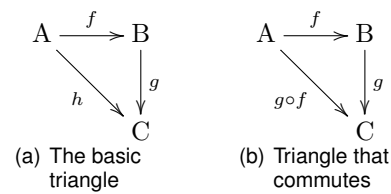

Objects and morphisms can be combined into diagrams, the simplest of which, and the most basic tool in category theory, is a triangle, as in Fig.~\ref{fig:triangles}.

\subsection{Commutativity, Identity, and Associativity}
\label{sec:CIA}
Here we introduce the notions  
required to ensure that a category is well defined. In section \ref{sec:VisProcCat} we see that these notions enforce some very basic conceptual hygiene on visualizations (such as the necessity of determining whether two versions of a visualization are `the same').

Given objects A, B, and C and morphisms $f:\A\longrightarrow \B$ and $g:\B\longrightarrow \C$, the first question a category theorist would ask about Fig.~\ref{fig:basicTriangle} is whether there is another morphism, $h:\A\longrightarrow \C$, that completes the triangle such that the result of applying $f$ then $g$ is the same as would be achieved by $h$. This morphism is the \emph{composition} of $f$ and $g$, or $g\circ f$. (Note the ordering: compositions are read right to left as the morphism $g$ is applied to the result of $f$.) If there is such a morphism $h$ then the triangle is said to commute, the objects and morphisms together form a category, and we can write the equation $h = g \circ f$.

Consider an informal example, the category of familial relations. Suppose A, B, and C are persons, $f$ is the mapping of `motherhood', and $g$ is `sisterhood'; then $g \circ f$ corresponds to `aunthood', the diagram commutes, and we have a category. But now suppose that $g$ is `friendship'. We do not have a well defined mapping for $g \circ f$ in this case (other than the tautological `the relationship I have with a friend of my mother'). Hence we cannot form a commuting triangle or create a category. To form a category encompassing both kinship and friendship we must enrich our vocabulary of morphisms to describe those relationships. Societies that are based on communal kin-groups rather than atomic families will tend to develop richer vocabularies to capture these composite relationships, hence their set of morphisms will tend to form well-defined categories. This is an example of how category theory can be used in practice: no one would suggest that kin-groups form vocabularies because of their category-theoretic properties, but the category-theoretic perspective suggests a set of questions that could be asked of that vocabulary, and a conceptual framework for analyzing it.

To form a category it is also necessary that all objects X each have a single identity morphism: $1_{\X}: \X\longrightarrow\X$ such that for every morphism $f:\A\longrightarrow\B$ we have $1_{\B}\circ f=f=f\circ 1_{\A}$ (i.e., mapping from A to B with $f$ has the same outcome before and after $1_{\A}$ and $1_{\B}$ are applied). We can view this diagrammatically: $$\xymatrix{
\A\ar@(ul,dl)[]|{1_{\A}}\ar[r]^{f}&\B\ar@(ur,dr)[]|{1_{\B}}
}$$
Thus, this notion of `sameness' is defined with respect to the morphisms in a category, rather than the identity of the object itself. The implications of this definition are explored in section \ref{sec:VisProcCat}.

The final requirement for a category to be valid is that morphism composition is associative, i.e., $(h \circ g) \circ  f = h \circ (g \circ f)$. In section \ref{sec:CTVis} we apply this definition of a category to the semiotic triad.

\subsection{Types of Morphism}
\label{sec:morphisms}
Here we introduce types of morphisms, an idea which lies at the heart of this work. Section \ref{sec:render} shows how these types correspond to, and have a direct impact upon, basic visualization concepts such as ambiguity, literalness, redundancy, and chart junk.

\begin{figure}[htbp]\centering\subfigcapskip=3pt
\subfigure[Monomorphism]{$\xymatrix@C=3.5em{\C\ar@<1.ex>[r]|{g_{1}}\ar@<-1.ex>[r]|{g_{2}}&\A\ar[r]^{f}&\B}$\label{fig:monic}}\quad\quad
\subfigure[Epimorphism]{$\xymatrix@C=3.5em{\A\ar[r]^{f}&\B\ar@<1.ex>[r]|{g_{1}}\ar@<-1.ex>[r]|{g_{2}}&\C}$\label{fig:epic}}\\
\subfigure[Endomorphism]{$\xymatrix@C=5em{\A\ar[r]^{f}&\A}$\label{fig:endic}}\quad\quad\quad\quad\quad
\subfigure[Isomorphism]{$\xymatrix@C=5em{\A\ar@<1.ex>[r]^{f}&\B\ar@<1.ex>[l]|{f^{-1}}}$\label{fig:isic}}
\label{fig:morphismTypes}\caption{Diagrams illustrating monomorphism, epimorphism, endomorphism, and isomorphism.}
\end{figure}
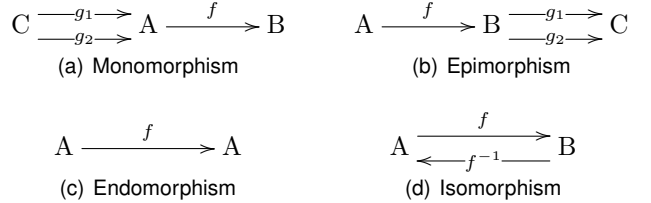

Morphisms are of several types, the main four which concern us here being monomorphic, epimorphic, endomorphic, and isomorphic. These are defined as follows.
\begin{itemize}
  \item $f:\A\longrightarrow\B$ is monomorphic (monic) iff $\forall g_{1}, g_{2}:\C\longrightarrow\A, g_{1}\neq g_{2} \implies f \circ g_{1} \neq f \circ g_{2}$. See Fig. \ref{fig:monic}.
    \item $f:\A\longrightarrow\B$ is epimorphic (epic) iff $\forall g_{1}, g_{2}:\B\longrightarrow\C, g_{1}\neq g_{2} \implies g_{1} \circ f \neq g_{2} \circ f$. See Fig. \ref{fig:epic}.\footnote{These definitions of monic and epic, through the use of the $\neq$ relational operator, rely on the closed world assumption. This is satisfactory for the classical logic of the category of sets but not for the intuitionistic logic of categories in general.}
  \item $f:\A\longrightarrow\A$ is endomorphic (endic) if it maps an object to itself. See Fig. \ref{fig:endic}.
  \item $f:\A\longrightarrow\B$ is isomorphic (isic) if there is an inverse mapping $f^{-1}:\B\longrightarrow\A$ such that $f \circ f^{-1}$ and $f^{-1} \circ f$ are both identity morphisms. That is, $f^{-1} \circ f = 1_{\A}$ and $f \circ f^{-1} = 1_{\B}$. See Fig. \ref{fig:isic}.
  \end{itemize}

Monomorphism, epimorphism, and isomorphism are generalizations of the more familiar set theoretic terms injection, surjection, and bijection, respectively. For example, let us define three set objects A, B, and C representing people and a morphism $f:\A\longrightarrow\B$ representing motherhood, such that $f$ maps each person in A to their respective mother in B. If $g_{1}$ and $g_{2}$ map people in one object to their friends in another then:
\begin{itemize}
  \item $f$ is \textbf{not} monic since two different friends of mine can have the same mother; that is, though we might have a pair of mappings such that $g_{1}\neq g_{2}$, the compositions $f\circ g_{1}$ and $f\circ g_{2}$ could be equivalent.
  \item $f$ \textbf{is} epic since two different friends of my mother must be different people (they cannot be the same person).
  \item $f$ is \textbf{not} endic since one cannot be one's own mother.
  \item $f$ is \textbf{not} an isomorphism since if $f^{-1}$ were `child' (the obvious candidate for an inverse map) then $f \circ f^{-1}$ could be my spouse (a person in B who is not my mother) and $f^{-1} \circ f$ could be my sibling (a person in A with whom I share a mother).
\end{itemize}

These simple examples illustrate how sensitive the properties of monomorphism and epimorphism are to the definitions employed for the underlying types. In section \ref{sec:application} we apply each of these definitions to the visualization category generated in section \ref{sec:CTVis}.

\section{Category Theory Applied to Visualization}
\label{sec:CTVis}
This section applies category theory to the semiotic triad. The end result can be seen in the commutative diagram in Fig.~\ref{fig:Process} but, as is usual with category theory, we proceed by constructing the diagram in stages, checking for commutativity at each step.

\begin{figure}[htbp]\centering
\subfigure[The Peircean semiotic triad]{
$\xymatrix{
\Obj\ar[r]^{f}\ar[dr]_{g \circ f}&\R\ar[d]^{g}\\
&\I\\\\
}$
\label{fig:peirceTriad}}\quad
\subfigure[Visualization]{
$\xymatrix{
\Data\ar[rr]^<<<<<<<<<{render}\ar[ddrr]|{understanding}&&\Representation\ar[dd]^{read}\\
\\
&&\Visualization
}$
\label{fig:visTriad}
}
\caption{The Peircean semiotic triad shown as commuting triangles}
\label{fig:triadCommute}
\end{figure}
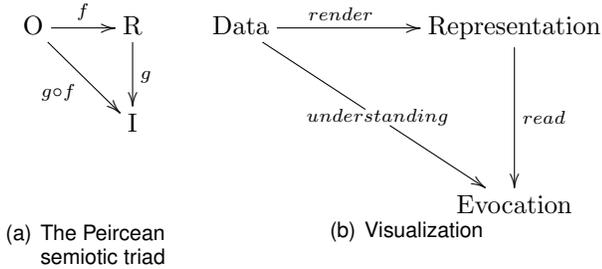

Peircean semiotics is based on a triadic relationship between object, representamen, and interpretant. 
We can draw our semiotic triad as the commutative diagram in Fig.~\ref{fig:peirceTriad} (where $\Obj$, $\R$, and $\I$ stand for object, representamen, and interpretant respectively).

In visualization the object is the data collected from a given system, the representamen is the representation, and the interpretant is the mental state evoked by the representation in the mind of the interpreter. Thus, we name the objects in our commutative diagram Data, Representation, and Evocation respectively. The morphisms between them we define as follows:
	\begin{itemize}
	\item The transformation from Data to Representation is named $render$ because the data are rendered in a given way so as  to represent the Data while maintaining structure and content. 
	\item The morphism between Representation and Evocation is called $read$ because the interpreter reads the Representation.
	\item To the composition $read \circ render$ we assign the name $understanding$ as a proper Representation leads to the reader understanding some aspect of the Data.
	\end{itemize}
	
Thus, Fig.~\ref{fig:visTriad} shows the Peircean semiotic triad presented as a commutative diagram which forms the core of the visualization process.

\begin{figure}[htbp]\centering
$\xymatrix@C=1.9em{
\System\ar[rr]^-{measure} &&{\Data}\ar[rrr]^{render}
\ar[ddrrr]|{understanding}
&&&{\Representation}\ar[dd]^{read}\\
\\
&&&&&\Visualization\\
}$
\caption{The objects System, Data, Representation, Evocation and the morphisms between them.}
\label{fig:CTInitial}
\end{figure}
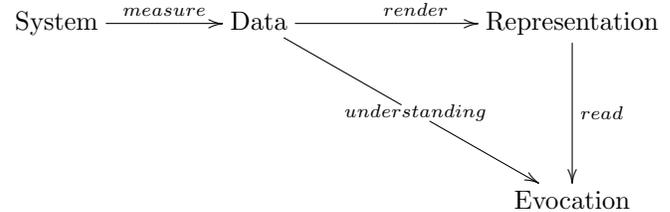

In visualization the object can be further decomposed into a set of data and a system which that data measures. Thus, the starting point for the visualization process is not data but the system from which the data were gathered. Furthermore, the interpretation of the representation leads not only to understanding the data but also to beliefs and inferences about the system, the truth of which can be tested. Thus, if we combine Figs \ref{fig:SonVis} and \ref{fig:visTriad} we get Fig.~\ref{fig:CTInitial}. A prototypical visualization process consists, then, of the following entities and processes, which we will describe using the category theoretic terms of objects and morphisms:
\begin{itemize}
	\item \textbf{System}: a real world system, object, or phenomenon, such as a class of students.
	\item \textbf{Data}: a set of data that describes some aspect of that System, produced by a \emph{measure}, such as test scores for those students.
	\item \textbf{Representation}: some visual, aural, haptic, or literal artefact of that Data, produced by a process of \emph{render}ing, such as a bar chart of their performance.
	\item \textbf{Evocation}: what the Representation evokes in the mind of the user through the user's reading of it, such as the teacher's viewing of the bar chart.
	\item And that Evocation is thus an understanding of the original data ($understanding=read\circ render$).
\end{itemize}
Fig.~\ref{fig:CTInitial} shows this expansion, though it should be noted that this diagram is incomplete and is to be read only as a stepping stone on the way to Fig.~\ref{fig:Process}. 

\subsection{The Visualization Process is a Category}
\label{sec:VisProcCat}
For these objects and morphisms to form a category certain conditions must apply.
\begin{enumerate}
	\item \textbf{Object Identity}: each of \{System, Data, Representation, Evocation\} must have an identity operation defined, and the kernel of this identity function will, in turn, identify an equivalence class of objects that are considered identical under this mapping. In particular this requires that we are able to decide unequivocally if two instances of the same System, Data, Representation, and Evocation are identical. This simple requirement forces a great deal of conceptual hygiene. Unless we can answer the following questions we are vulnerable to the accusation that the visualization process is not well-defined.
	\begin{enumerate}[\IEEEsetlabelwidth{4)}]
		\item $1_{\System}$: The problem of identifying systems is a common and urgent one in most empirical science; we cannot talk about replicating results unless we are doing the same things to the same systems. Moreover, unlike a data set which captures a snapshot of a system and is static, a System may experience change over time. For example, in a System of a class of students the individual students age and mature, change their clothes, and some may even drop out of the course. How may we know when two System objects have the same denotation? Agreed criteria are needed for establishing System identity which will allow a System to experience change over time while still being considered to be the same System.
		\item $1_{\Data}$: When do we say that two sets of data, such as test scores, are the same? Do the absolute scores matter or is it the same set of scores when expressed as a percentage? What degree of precision is required? If we had a very large set of scores (such as when studying the changes in national exam performance) then identity might be defined in terms of there being no statistically significant difference between samples or aggregate distributions.
		\item $1_{\Representation}$: When are two representations the same? Is a printed version the same representation as an on-screen version, or an auditory display equivalent to a visual one? 
		\item $1_{\Visualization}$: What does it mean to say that two users form the same mental picture of the data? The obvious problems in determining internal psychological states mean that this issue is usually operationalized in terms of the ability to answer questions about the data (`which student did best?', `has average performance gone up or down?'), where the same answer implies the same understanding.
	\end{enumerate}
	\item\textbf{Morphisms must be Maps}: recall there can be only one object at each end of an arrow. In the case of $render$ this requires that a single set of Data (to within $1_{\Data}$) generates a single Representation. This is not obviously true. For example, visualization tools that allow interactive data exploration seem capable of generating many representations of a single data set. But in this case it is the tool, bound to that Data, that is considered to be the Representation rather than any particular state or view that it produces.
	\item\textbf{Commutativity of Morphisms}: As section \ref{sec:CIA} showed, commutativity is required. For example, in Fig. \ref{fig:CTInitial} the result of \emph{read}ing the Representation produced by \emph{render}ing the Data is an \emph{understanding} of the Data. If the Data is \emph{render}ed and then \emph{read}, but as a result the reader does not understand the data then the process of visualization has failed.
	\item\textbf{Morphism Associativity}: As per section \ref{sec:CIA}, the composition of morphisms must be associative. 
\end{enumerate}
Visualization processes for which conditions 1--4 are  satisfied can be considered as valid categories satisfying the axioms of category theory.

\subsection{The Intension of the Visualization Process}
\label{sec:intension}
Fig.~\ref{fig:CTInitial} shows the visualization process as a category, but this represents a single concrete instance of this process. It could represent, for example, the production and consumption of a single particular scatter plot from a single particular set of data. We now need to generalize this notion to describe, for example, the properties of scatter plots in general. In this section we generalize each of the objects of the visualization category.

\subsubsection{Schema is the Generalization of Data}
In the case of the Data we have the familiar notion of a Schema, which refers to the structure rather than any particular values of the data. For example, a relational database of tables and attributes is defined using a database schema, which is then filled with values during its lifetime of use. The Schema is the intension (the constant conditions that capture the set of all possible values for that data set) while the Data is the extension (the variable set of actual values).\footnote{In philosophy a distinction is drawn between a term's intrinsic meaning or its \emph{intension} and its denotation, or \emph{extension}. Frege gave the example of the \emph{morning star} and \emph{evening star}. The two terms have different meanings (intensions) but both have the same denotation (extension), the planet Venus \cite{Sowa:2000}.} Not all Data has a corresponding Schema (generally known as unstructured data). Examples include natural language text which can be represented, for example, by pictorial illustrations. However there can be no rules governing the layout of this Representation: there may be a set of Representations of a similar style (for example, several illustrations in a single book) but each is individually inspired by the text it is designed to Represent. Unless the Data can be generalized into a Schema, there can be no corresponding generalization of the Representation into a Layout. This criterion also helps to distinguish between works that are data art or data-driven art (see section \ref{sec:infograph}).

\subsubsection{Layout is the Generalization of Representation}
We can also generalize the notion of a particular Representation to that of a Layout, that is, the way in which Data belonging to a Schema are represented. For example, the two scatter plots in Fig.~\ref{fig:PlainCharts} are different Representations, but share a Layout. Layout is a familiar notion from data visualization tools (such as the charting feature in spreadsheet programs) that allow the user to choose which of many options are used to graphically represent selected data. The tool uses a set of $rules$ to map from a data Schema to a representational Layout.  

\subsubsection{Questions are the Generalization of Evocation}
\emph{Generalization} is the process of splitting the extension into an unsaturated (incomplete) and a saturated (complete) part (in Frege's sense of the terms \cite{Frege:1891}). It is the former that constitutes the intension. Data, for example, may be associated with a Schema, such as a table, and the values that can fill the empty spaces. Representations can be split into a Layout, such as a set of axes, and the marks that are placed in the space defined by the axes. In the case of the contents of the mental states evoked by \emph{reading} a Representation, the equivalent is to split the proposition describing that mental state into a property for which there may be some object of which it can be truthfully predicated. That is, if \emph{reading} a Representation, such as a bar chart of exam results, evokes the thought that `Alan got the best mark' (or $best\_mark(Alan)$), then the generalization of this is the question `Who got the best mark?' (or $best\_mark(\_)$).
That is, Questions are system predicates at the intensional level while Evocation involves the extension of those predicates for specific cases. Evaluation of these extensional predicates allows truth statements about the System to be tested.

Completing the \emph{generalization} mapping, from extension to intension, provides a salutary reminder of the importance of the underlying Question in visualization; ``From Killer Questions Come Powerful Visualizations'' as Johnstone puts it \cite{Johnstone:2011}. In software development terms, answering a question is the \emph{requirement} of the Representation.
This is true even in the case of \emph{exploratory} data analysis (the terminology is due to Tukey \cite{Tukey:1977}), where the purpose is not to answer a specific prior Question or hypothesis, but to discover hypotheses worth subsequent testing using conventional \emph{confirmatory} data analysis. Exploratory data analysis is what happens when you don't know what question you're trying to answer. Confirmatory data analysis starts with Data and a Question and then seeks an answer using a Representation. Exploratory data analysis starts with Data and a Representation, and then seeks a Question worth asking.

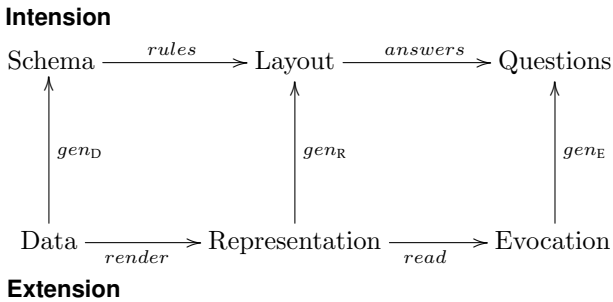
\begin{figure}[htbp]
  \centering
  \scalebox{1.0}{$$\xymatrix@-3mm{
\put(1,20){\txt{\bfseries\sffamily\small Intension}}
\Schema\ar[rr]^{rules}&&\Layout\ar[rr]^{answers}&&\Question\\
\\
 \Data\ar[uu]_{gen_{\text{D}}}\ar[rr]_>>>>>>>>>{render}&&\Representation\ar[uu]_{gen_{\text{R}}}\ar[rr]_{read}&&\Visualization\ar[uu]_{gen_{\text{E}}}
\put(-230,-15){\txt{\small\sffamily\bfseries Extension}}
}$$}
        \caption{Introducing the intensional level of the category generalizing the extension with $gen_{D}:\Data\longrightarrow\Schema, gen_R:\Representation\longrightarrow\Layout, 
        gen_E:\Visualization\longrightarrow\Question$ }
    \label{fig:DSRL}
\end{figure}

Fig.~\ref{fig:DSRL} shows the Data, Representation, and Evocation objects from our visualization category at the extensional level and Schema, Layout, and Questions as their respective  generalizations, or intensions.

\subsection{Completion of the Visualization Category}
\label{sec:closure}
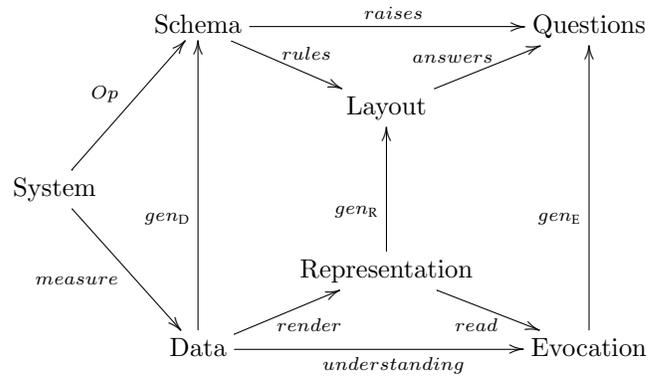
\begin{figure}[htbp]
$$\xymatrix@C=0.5em@R=1.6em{
&&\Schema\ar[rrrr]^{raises}\ar[drr]^{rules} &&&&\Question\\
&&&&\Layout\ar[urr]^-<<<<<<{answers\   }\\
\System\ar[uurr]^{Op}\ar[ddrr]_{measure}\\
&&&&\Representation\ar[drr]_{read}\ar[uu]^<<<<<<{gen_{\text{R}}}\\
&&\Data\ar[uuuu]^<<<<<<<<<<<<<<<{gen_{\text{D}}}\ar[urr]_{render}\ar[rrrr]_{understanding} &&&& \Visualization\ar[uuuu]^<<<<<<<<<<<<<<<{gen_{\text{E}}}\\
}$$
\caption{Intermediate mathematically closed visualization category with terminal object Questions.}
\label{fig:CTPartial}
\end{figure}

Combining our partial diagrams (Figs. \ref{fig:CTInitial} and \ref{fig:DSRL}) we get the diagram in Fig.~\ref{fig:CTPartial}.
From a category-theoretic point of view, that is, by considering its formal structural properties, this diagram is now complete as it is closed in the mathematical sense.  Seeking closure requires us to find a terminal object, an object at which all paths of possible morphism compositions terminate. Although Fig. \ref{fig:CTPartial} does have such an object (Questions), we would not say that this is the terminal object
of the visualization process. Therefore, we need to identify a new terminal object that reflects the true end of this process.

\subsubsection{Knowledge is the Terminal Object of Visualization}
Data is not the start of the visualization process, and nor is the ability to answer Questions about that Data the end. What we are seeking is Knowledge of the System. Visualization starts with a System that we measure in various ways to generate Data. That Data will always be partial (in both senses of the word) but it is all we have. Similarly, the only Knowledge we can gain is that which can be deduced from the evidence presented in the Representation, and the Questions define what Knowledge we can gain from the visualization process. That is, Knowledge operationalizes the Questions by allowing the abstract nature of the Questions to be practically measured or assessed. It is by answering questions that one gains knowledge. This relationship is also analytic: Knowledge is not well-defined unless it is capable of answering questions, and those questions are  epistemologically prior to the knowledge of the answers. (One can imagine a question to which there is no answer, but not an answer for which there is no question.)

The ultimate goal of the visualization process is to gain Knowledge of the original System. When this succeeds (when the diagram commutes) the result is a \emph{truth} relationship between the Knowledge and the System. When this process breaks down and we fail to deduce correct conclusions then the diagram does not commute.\footnote{Note, that although this is a strongly realist and representational use of terms such as truth and knowledge it does not necessarily imply a commitment to objectivism about the status of that knowledge. See, for example, Faith \cite{Faith:2000}.}
\subsubsection{The Completed Category}
\begin{figure}[htbp]\small
$\xymatrix@C=-0.1em@R=1.8em{
&&\Schema\ar[rrrr]^{raises}\ar[drr]^{rules} &&&&\Question\ar[ddrr]^{Op}\\
&&&&\Layout\ar[urr]^-<<<<<<{answers\   }\\
\System\ar[uurr]^{Op}\ar[ddrr]_{measure}\ar[rrrrrrrr]^<<<<<<<<<<<<<<<<<<<<<<<<<<<<<<<<<<<<<<{truth}|!{[rr]}\hole|!{[rrrr]}\hole|!{[rrrrrr]}\hole &&&&&&&& \Knowledge\\
&&&&\Representation\ar[drr]_{read}\ar[uu]^<<<<<<{gen_{\text{R}}}\\
&&\Data\ar[uuuu]^<<<<<<<<<<<<<<<{gen_{\text{D}}}\ar[urr]_{render}\ar[rrrr]_{understanding} &&&& \Visualization\ar[uurr]_{infers}\ar[uuuu]^<<<<<<<<<<<<<<<{gen_{\text{E}}}\\
}
$
\caption{The visualization process as a category with terminal object Knowledge and initial object System.}
\label{fig:Process}
\end{figure}
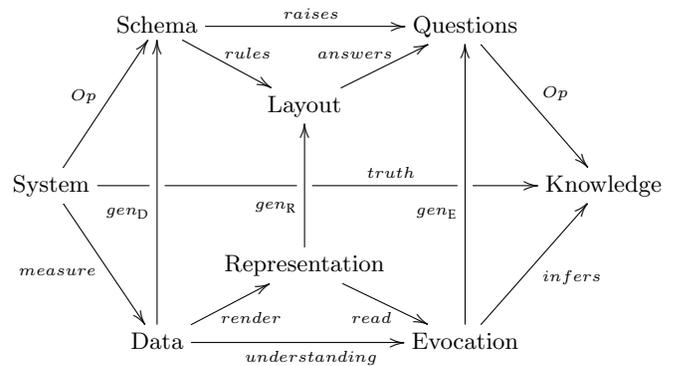

The full diagram describing the general visualization process is given as Fig.~\ref{fig:Process} and a specific example  is shown in Table ~\ref{tab:ExampleProcess}. In this example we start with students in a mathematics class (System). We want to know how well they are performing, overall and individually (Knowledge), so we determine that we must gather data about their performance in a test (Schema). This Schema, as well as being a generalization of the Data is also a time-invariant $descriptive$ abstraction of the System.\footnote{Formally, a Schema has ``the structure of a continuant which does not specify time or timelike relationships'' \cite[p. 73]{Sowa:2000}.} Having gathered the results (Data) we choose a bar chart (Layout) that will be capable of showing the distribution of overall and individual marks (Questions). The Schema acts as a frame and determines, or $raises$, what Questions can be answered by the Layout. For example, from a table with two columns labelled `Student' and `Grade' we can tell that the Data denoted by this Schema will let us answer questions like ``What was the average mark?'', but not questions like ``Who was the tallest?''.

The Layout, in turn, is used to $answer$ the Questions, but only some Schema-derived Layouts are capable of doing this. For example, a bar chart will enable us to spot the best student but a pie chart would not. 

Showing the data (Representation) in this way evokes an understanding of the data that the cohort achieved an average mark of more than 70, that Alan got the highest mark, etc. So now we have Knowledge about our System (the Maths class): we know who performed best of all, that nobody failed, that the average mark was satisfactory (this requires prior context), etc.
\begin{table}[htbp]\centering
\renewcommand{\arraystretch}{1.3}
\caption{Example of how the visualization category applies to objects in practice}
\label{tab:ExampleProcess}
\begin{tabular}{cp{5cm}}
\hline
\textbf{Category Object} & \textbf{Example} \\
\hline
System & A cohort of students on a course \\
\hline
Schema & \parbox[c]{1em}{\includegraphics{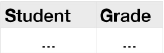}} \\ \hline
Data & \parbox[c]{1em}{\includegraphics{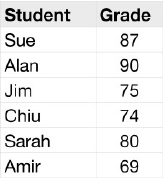}} \\ \hline
Layout & \parbox[c]{1em}{\includegraphics{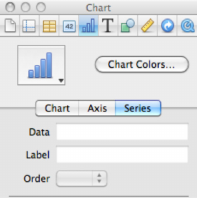}} \\ \hline
Representation & \parbox[c]{1em}{\includegraphics{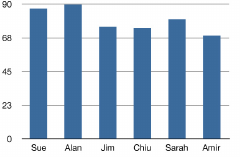}} \\ \hline
\multirow{2}{*}{Questions} &$best\_mark(\_)$ \\ 
&$average\_mark(\_)$\\ \hline
\multirow{2}{*}{Evocation} &$best\_mark(Alan)$\\&$average\_mark(> 70)$ \\ \hline
\multirow{3}{*}{Knowledge} & ``Alan performed the best'',\\
 & ``Nobody failed'', \\&``The average mark was satisfactory''\\
 \hline
\end{tabular}
\end{table}

\section{Results: Applying the Category}
\label{sec:application}
With this category corresponding to the visualization process we can use the concepts of category theory to consider its properties, and show how common intuitions about visualization can be defined more formally.
\subsection{Properties of the \emph{render} Morphism}\label{sec:render}
\label{sec:morphism}To illustrate how this may be done, we will consider what it means in visualization terms, for the $render$ morphism to be monomorphic, epimorphic, isomorphic, and endomorphic. In each case we take a standard  category theoretic definition and apply it to the visualization process category and we find that this yields a property or issue that is important for visualization. 

\subsubsection{Monomorphism Corresponds to Sensitivity}
A \emph{render} morphism is monic iff $\forall measure_1, measure_2: \System\longrightarrow\Data:
render\circ measure_{1}\neq render\circ measure_{2} \implies measure_{1}\neq measure_{2}$.

That is, if the System is measured in two different ways the resulting Representations will necessarily be different.  
For example, suppose we assess student performance with two different tests the results of which are captured by $measure_{1}$ and $measure_{2}$ and represent them in two ways: a simple textual description (e.g., ``John was the best student''), and a bar chart. The differences in the test would not make any difference to the textual description but they would to the bar chart (assuming that identity morphisms on the Representations and Data are well-defined). We would  normally describe this in terms of the sensitivity of the visualization. Sensitivity is usually assumed to be a desirable property of a representation (and this assumption may often be valid) but the point here is to show that this important property may be defined formally and precisely. 

\subsubsection{Epimorphism Corresponds to Non-Redundancy} 
A \emph{render} morphism is epimorphic iff  $\forall read_1, read_2:\Representation\longrightarrow\Visualization: 
read_{1}\circ render \neq read_{2}\circ render \implies read_{1}\neq read_{2}$.

\begin{figure}[htbp]
  \centering
    \mbox{
     \subfigure[The data set]{\includegraphics{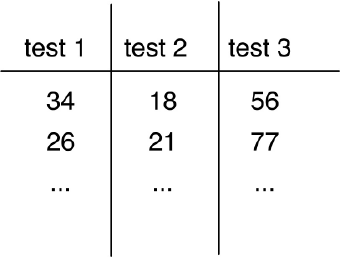}
   \label{fig:redundancyA}} \quad
      \subfigure[A redundant representation: the \emph{test 3} dimension is mapped to both shade and size]{\includegraphics{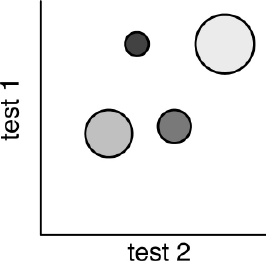}
	\label{fig:redundancyB}} 
      }
    \caption{Redundancy in representations}
    \label{fig:redundancy}
\end{figure}

That is, if two people \emph{read} the same Representation in different ways they will reach different \emph{understandings} of the Data. Although this may seem tautological, there are important cases when it is not true. Consider a set of data in which three attributes are measured for each sample (e.g., student performance on three different tests). This data could be represented using a conventional scatter plot in which the $x$-coordinate corresponds to test 1, $y$ to test 2, and both the size and shade of each point correspond to test 3 (see Fig.~\ref{fig:redundancy}). One individual may notice the position and size of each point, whereas another may notice the shade. They would draw identical conclusions about the data, but they have read the representation in different ways. The representation in this case is redundant as test 3 is represented in two different ways; hence there is more than one way of gaining an understanding of that information from a single representation. 

\subsubsection{Endomorphism Corresponds to Literalness}
A \emph{render} morphism is endomorphic (a mapping of an object back onto itself) iff $Data=Representation$. That is, some Representations consist in simply presenting the Data (for example, in a spreadsheet or printed table). These, too, are valid visualizations, and the concept they demonstrate is \emph{literalness}.\footnote{N.b., we are bracketing the discussion  over the metaphysical status of logical data compared with that of a physical representation.}  

Contrast this with Cox's notion that visualizations are necessarily metaphorical \cite{Cox:2006}, that there is a direct relationship between visualization and the mapping process (cognitive and creative) in metaphor theory. She says (p. 89):
\begin{quote}
Linguistic and visual metaphors are defined as mappings from one domain of information (the source) into another domain (the target). Likewise, data-viz maps numbers into pictures, resulting in visaphors, digital visual metaphors. 
\end{quote}
However, we can see that endomorphic $render$ings are valid visualizations that are not metaphorical.

\subsubsection{Isomorphism Corresponds to Non-Ambiguity}\label{sec:isic}
A \emph{render} morphism is isomorphic iff there is an inverse morphism (which we will call \emph{decode}) such that 
$decode\circ render=1_{\Data}$ and $render\circ decode=1_{\Representation}$.
Thus, the \emph{decode} morphism lets us recover the original Data from a Representation \emph{render}ed from it (to within the degree of accuracy determined by $1_{\Data}$). Non-isomorphic Renderings are thus ambiguous in the sense that two different data sets may generate the same Representation. Sometimes this is a valuable property, for example in representations that aggregate or filter large or complex data sets into simpler forms. In other situations it is less desirable. 
Further, as isomorphism implies both monomorphism and epimorphism, a rendering that is isomorphic, in addition to being non-ambiguous, will also exhibit sensitivity and an absence of redundancy. These properties are summarized in Table \ref{tab:morphSummary} which shows what combinations of morphism properties determine what characteristics of a particular $render$ing (e.g., only endic $render$ings are literal, monic ones exhibit sensitivity, non-isomorphism leads to ambiguity, etc).

\begin{table}[htbp]\centering
\renewcommand{\arraystretch}{1.3}
\caption{Summary of morphism properties as applied to the $render$ morphism.}
\label{tab:morphSummary}
\begin{tabular}{llll|lll}
\hline
&\multicolumn{3}{c|}{\bfseries Non-endomorphic}&\multicolumn{3}{c}{\bfseries Endomorphic}\\
&\textbf{Monic}&\textbf{Epic}&\textbf{Isic}&\textbf{Monic}&\textbf{Epic}&\textbf{Isic}\\
\hline
\textbf{Sensitive} &\checkmark&&\checkmark&\checkmark&&\checkmark\\
\textbf{Non-redundant} &&\checkmark&\checkmark&&\checkmark&\checkmark\\
\textbf{Non-ambiguous} &&&\checkmark&&&\checkmark\\
\textbf{Literal} &&&&\checkmark&\checkmark&\checkmark\\
\hline
\end{tabular}
\end{table}

\subsection{Non-Epimorphic Layout Implies Chart Junk}\label{sec:nonepic}
In this section we show how one common contentious issue in information visualization (chart junk) can be characterized using category theory.
\begin{figure}[htbp]
  \centering
     \subfigure[Plain charts]{\includegraphics{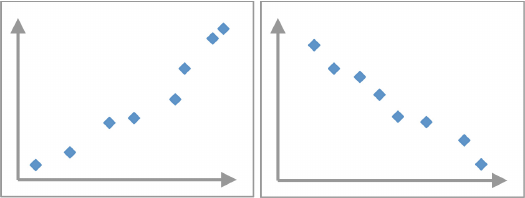}
       \label{fig:PlainCharts}}\\
      \subfigure[Arbitrary chart junk]{\includegraphics{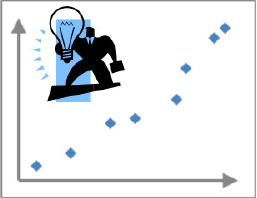}
	\label{fig:ChartJunkArbitrary}}\\
      \subfigure[Schema-derived chart junk]{\includegraphics{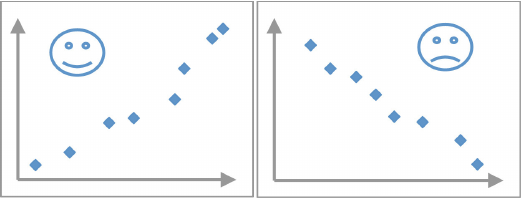}
        \label{fig:ChartJunkSchema}}
    \caption{Chart Junk}
    \label{fig:ChartJunk}
\end{figure}
In the scatter plots in Fig.~\ref{fig:PlainCharts} every element of the Layout was derived from some element of the Schema. If we write these objects as sets, then we can say that the morphism between them is a surjection:
\begin{eqnarray*}
\Schema &=& \{Instances, Attributes\}\\
\Layout &=& \{Points, Axes\}\\
rules(Instances) &=& Points\\
rules(Attributes) &=& Axes
\end{eqnarray*}
However, this is not always the case. For example, suppose we decorate one of our scatter plots with a figure as in Fig.~\ref{fig:ChartJunkArbitrary}. What is the Layout in this case? If the Layout is a \emph{generalization} of this particular Representation then it will include the decoration. But this decoration is not derived epimorphically using a \emph{rule} from any part of the Schema --- it is an arbitrary addition (perhaps inspired from some property of the System not captured in the Data). It is `chart junk'.

The following aspects of this definition of chart junk should be noted.
\begin{enumerate}
	\item This use of the term is not intended to be pejorative: chart junk can be useful in aiding understanding of the system \cite{Bateman:2010}. The purpose of this category-theoretic definition is to highlight the difference between decorative elements in Layouts that communicate Data (those that are derived from a Schema) and those that do not. 
	\item It is a much narrower use of the term than Tufte's original definition \cite{Tufte:1983} which defined chart junk as all unnecessary, redundant, or non-data ink. Consider the example in Fig.~\ref{fig:ChartJunkSchema} in which scatter plots are decorated with faces indicating the data trend. The decoration is redundant chart junk in Tufte's sense, but not in ours since it communicates something about the data. There is a rule for deriving this element of the Layout from the Schema which thus ensures that Fig.~\ref{fig:Process} commutes. 
	\item There is a difference between redundancy at the extensional level of the Data and Representation introduced in Fig.~\ref{fig:DSRL} and arbitrary chart junk at the intensional level of the Schema and Layout. A Representation may include some redundancy even though there is no chart junk. In the example of the scatter plot in which a single Data attribute is represented using two retinal attributes such as in Fig.~\ref{fig:redundancyB} --- or, indeed, the example in Fig.~\ref{fig:ChartJunkSchema} of a scatter plot decorated with a face representing the polarity of the correlation --- we have redundancy in the Representation, but the Layout \emph{rules} are surjective (epic): every element of the Layout is the product of some aspect of the Schema. Redundancy at the level of Data and Representation may not be arbitrary at the level of Schema and Layout. Conversely, chart junk is not (necessarily) redundant, it is arbitrary.
\end{enumerate}

\subsection{Intensionality and Infographics}
\label{sec:infograph}
\begin{figure}[htbp]
\centering
\includegraphics{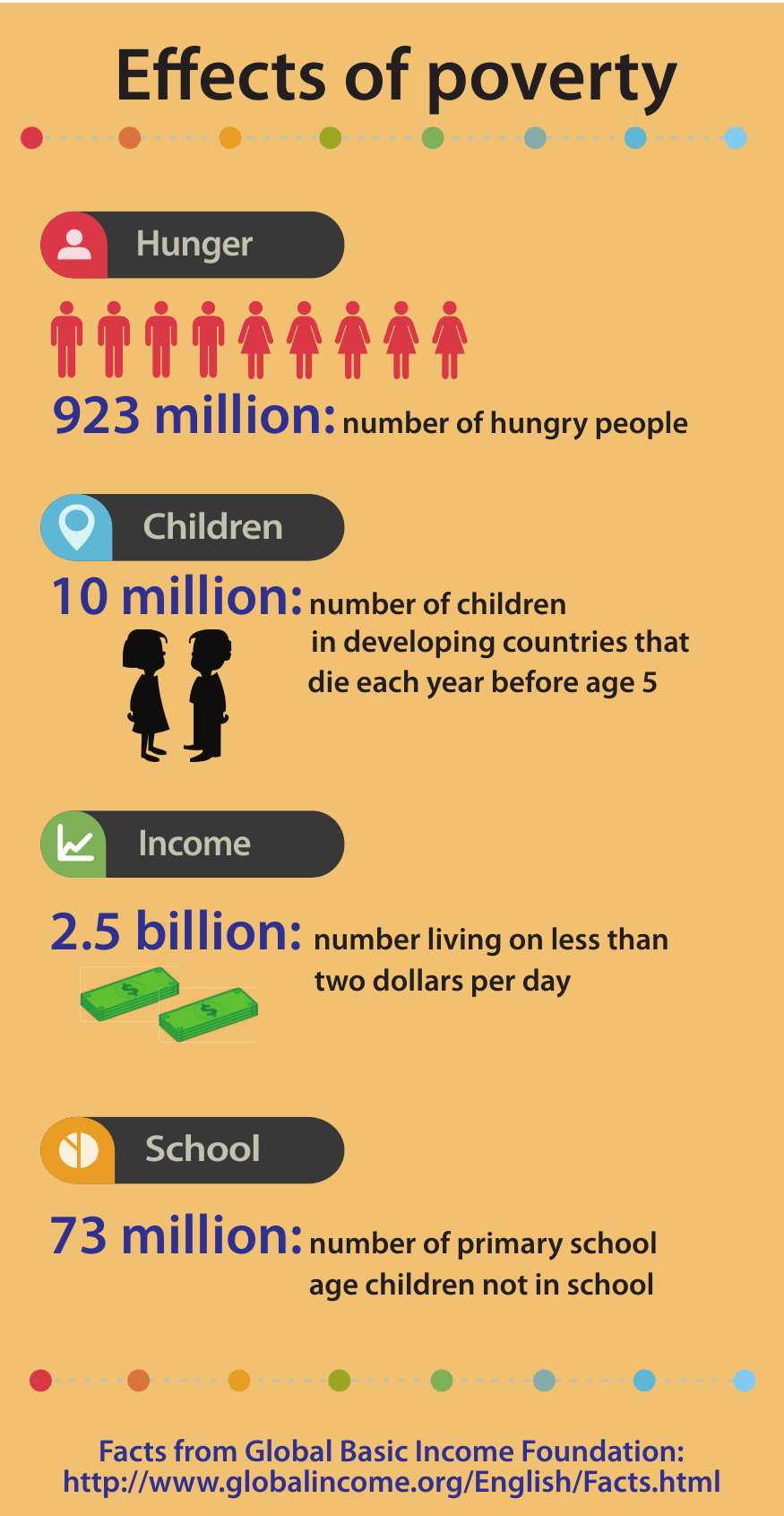}
\caption{Infographic showing facts about world poverty. Facts taken from the Global Basic Income Foundation at \protect\url{http://www.globalincome.org/English/Facts.html}}
\label{fig:infographic}
\end{figure}
\begin{figure*}[htbp]\centering
    \subfigure[Exploded `Support' section]{
        \includegraphics{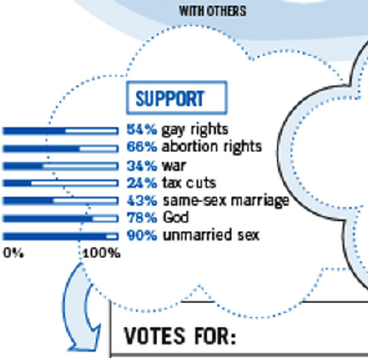}
   \label{fig:lrsupport}} 
      \subfigure[Full chart showing both sides]{
       \includegraphics{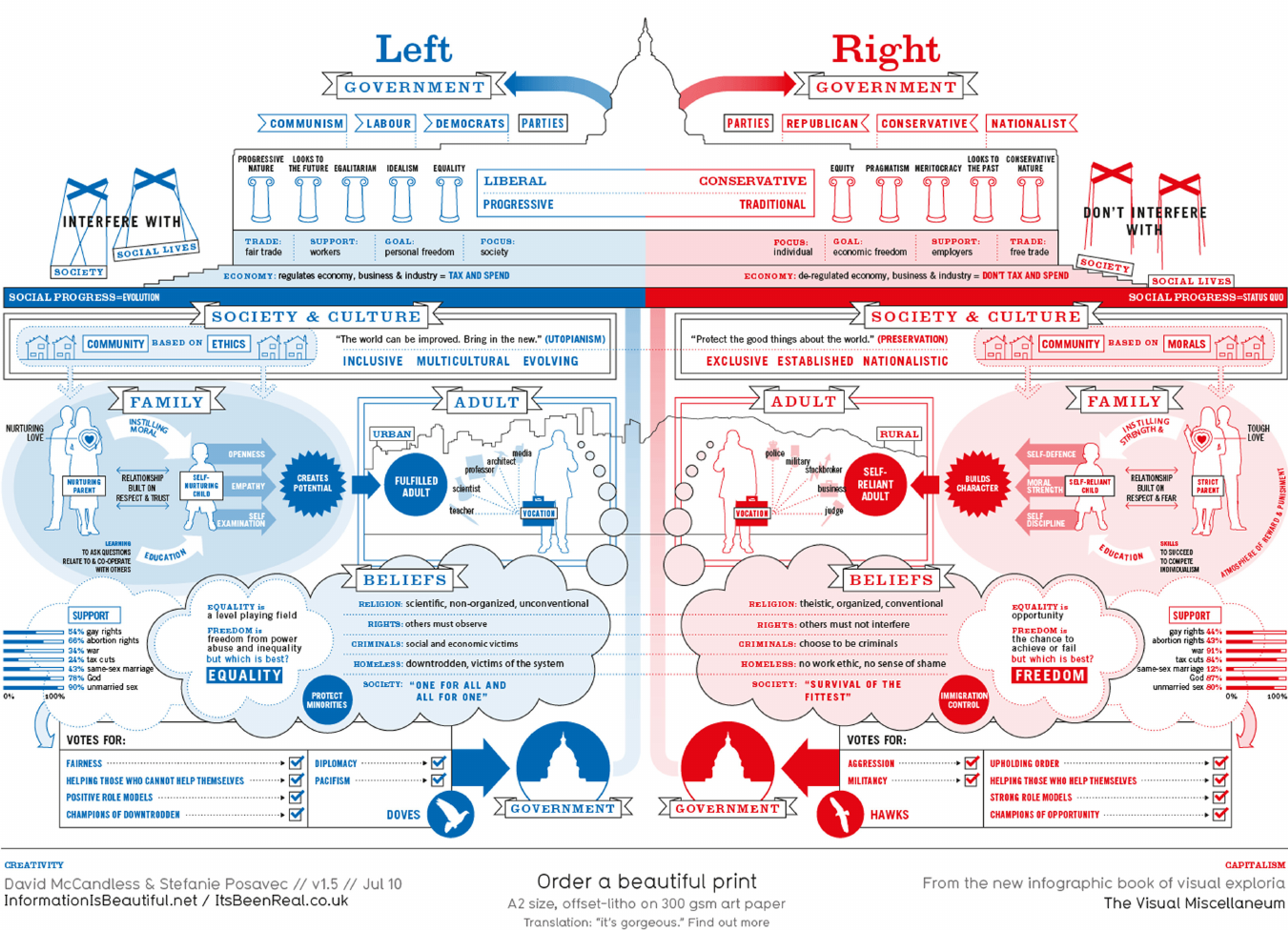}
      \label{fig:lrfull}}
    \caption{Infographic illustrating political differences. David McCandless 2009 --- \protect\url{http://www.informationisbeautiful.net/visualizations/left-vs-right-us/}}
    \label{fig:leftright}
\end{figure*}

For an example of the power of intensional generalization, compare the two infographics in Figs. \ref{fig:infographic} and \ref{fig:leftright}. In Fig. \ref{fig:lrfull} both sides of the diagram follow the same pattern; this pattern is a schema. The existence of this intensional generalization means the representation can support two operations that Fig. \ref{fig:infographic} cannot. The first operation is that it can generate questions that it can then answer (for example: ``I see that 54\% of left-leaning people support gay rights; so what proportion of right-leaning individuals do?'' --- Fig. \ref{fig:lrsupport}). The second operation that Fig. \ref{fig:lrfull} can support is that the layout can be generalized to more examples (e.g., how would this diagram apply to European socially progressive and economically libertarian political parties?). Having no intensional level, the example in Fig. \ref{fig:infographic}  can support neither of these operations. It communicates information that evokes understanding, but does not empower the user to interrogate or reuse the information it presents.

\section{Related theory}
\label{sec:theory}
\subsection{Algebraic Semiotics}
Both Peirce and Saussure understand signs in terms of relationships and mappings between signs and sign systems. Thus, they seem natural candidates for the category theory treatment. The first effort to apply category theory to semiotics was that of Goguen and Harrel \cite{Goguen:2005} who attempted a formalist treatment of the semiotics of information visualization and user interface design, which they described as `algebraic semiotics'.

The objects in Goguen's algebraic semiotics are sign systems. The actual definition is complex, involving theoretical apparatus taken from mathematical algebra but, the key elements are signs, constructors, and axioms. The signs form the vocabulary, or set of all possible signs. The constructors provide a systematic way of generating those signs. And the axioms constrain those signs. Consider Goguen's example of a simple time of day system which shows the number of minutes since midnight. The signs are the set of natural numbers generated using two constructors: the constant 0 (representing midnight) and a successor operation, $s$, where for a time $t$, $s(t)$ is the next minute. A single axiom, $s(1439)=0$, constrains the set of generated signs to a 24 hour day.
Goguen thus defines signs in terms of generalized structures from which the validity of any particular sign is derived. In terms of of first-order logic, the sign system is a theory of which any particular sign is a model.

Goguen then considers the mappings between sign systems, which he calls semiotic morphisms. Consider a slow, regular sand glass containing 1440 grains of sand in which one grain of sand falls every minute, and which is turned when the last grain falls. We can define a mapping from the time of day system by respectively mapping its elements (constant 0, the $s$ constructor, the set of numbers, the $s(1439)=0$ axiom) to our new elements (empty lower glass, the falling of a grain of sand, the possible piles of sand, turning the glass).

Goguen's is a strongly structuralist theory in two senses. First, signs are defined as such in virtue of their membership of, and role within, a sign system. It is the structure of the sign system that defines its constituents as signs. Second, the only relationships considered --- the semiotic morphisms --- are between sign systems rather than between sign systems and either external or mental states. It all happens within the `third order' \cite{Deleuze:2002}. 
In particular, there is no distinction in this framework between a set of data and a visualization of that data. Data is just another sign system.

We find this structuralism problematic. For example, how can this framework be used to discuss the quality of a visualization? Goguen attempts an answer to this `narrowly' by characterizing how well his semiotic morphisms preserve the structure of a sign system. We argue that the solution, as presented in this article, is a post-structuralist scheme that expands the use of category theory to explicitly incorporate other elements of the visualization process, including the visualization's context, and how the visualization is used in practice.

\subsection{Non-Category Theoretic Approaches}
Our framework is method-blind in that it does not specify \emph{how} the various objects (Schema, Data, Questions, etc.) and morphisms ($render$, $raises$, etc.) should be defined. Other formal approaches to visualization have sought more to explore the \emph{hows} by understanding the biology, as it were, of different visualization techniques. They focus variously on classification (taxonomy), on describing the differences and commonalities between visualization techniques, on codifying graphic design and layout rules, and on understanding the different representational processes employed in visualization of all kinds.

Tory and M\"oller \cite{Tory:2004} show how theoretical models and taxonomies focus on understanding the design space and design choices of the visualization process. Their approach seeks to understand and clarify the differences between different types of visualization to show how ``traditional divisions (e.g., information and scientific visualization) relate and overlap'' \cite{Tory:2004}. The focus is on the process level letting us see both the relationships between different task types and what types of task each design model enables users to perform.

Continuing in the taxonomic vein Chi \cite{Chi:2000} aimed to understand how visualization techniques work and how they fit into different design spaces. Understanding the different operating steps and how they are re-used in different visualization techniques enables, it is claimed, more rapid implementation of visualizations. 

We see from Fig. \ref{fig:Process} that the taxonomic techniques above deal with, and could thus be used in the specification of the Schema and Layout objects and the $rules$, and $render$ing morphisms. Consider Tory and M\"oller's taxonomy which suggests suitable schemata for different data types. Two-dimensional data with one dependent and one independent variable, for instance, can be represented using scatter plots and bar charts whilst data with any number of dependent and independent variables can be mapped to charts with colour, multiple views, glyphs, and parallel co-ordinate plots. Chi's taxonomy gives a very detailed analysis of 36 visualization techniques (Representation + Layout) showing what kinds of data they are well-suited to. Thus, taxonomy-based treatments help the visualization designer with specifying the intensional level of the category for a given System and Data.

Mackinlay \cite{Mackinlay:1986} analyzed information graphics and combined first-order logic  with a `composition algebra' to codify graphic design criteria to allow construction of a tool that automatically designs graphical presentations of relational information. The composition algebra's rules enable multiple data sets with common axis dimensions to be displayed on a single combined plot. It is very much concerned with layout rules and representational techniques. Thus, we can see that this technique might help in determining the $rules$ morphism used in the specification of Layout. Mackinlay offers `expressiveness criteria' that indicate the encoding technique that should work for each data mapping. For example, a Layout may use size as a display attribute effectively for ordinal or quantitative data while colour, orientation, and shape will all work well for nominal data. Card and Macklinay \cite{Card:1997} took this foundation further and organized the visualization design space as a framework that can be used to design new or augment existing visualizations. Principally, their work is concerned with the intensional decisions of how best to map a given Schema to a Layout given the information requirements (Questions) of the task.

Perhaps closest in scope to our work, Ziemkiewicz and Kosara \cite{Ziemkiewicz:2009} used set theory to explore
the mapping between data and representation. They asserted that for a visualization to be readable the rendering must be both injective and surjective, that is, bijective (isomorphic in our category). Non-injective mappings, they say, can lead to representations in which more than one data item is mapped to the same visual element leading to ambiguity (see section \ref{sec:isic}) while non-surjective mappings lead to representations containing marks not derived from the data (which would include chart junk, see section \ref{sec:nonepic}). Their analysis seems to preclude chart junk as a valid element in a visualization, while our scheme makes no assessment of the desirability of chart junk.  Ziemkiewicz and Kosara add a further stipulation that all information visualizations must involve non-trivial interactivity and be syntactically notational. Using these three criteria they then classify and distinguish between different types of visualization (e.g., scientific visualization, information graphics, etc.). Within information visualization they identify three classification axes: linear vs non-linear mappings, how information loss is treated in the rendering, and the semantic notationality of the system. This theoretical framework enables taxonomic analysis within information visualization and between different non-information visualization techniques. With the focus being on the properties of the representational mapping their work sits, again, in the intensional part of our category centered on the $rules$ morphism (the intension of $render$).

Ziemkiewicz and Kosara's work is very helpful in understanding different visualization processes in a taxonomic sense (though they tend to exclude narrative visualizations for their lack of interactivity). The set theoretic approach allows sensitivity, redundancy, and ambiguity to be  defined in a similar way to our approach but category theory enables us to go further and differentiate between:
 \begin{enumerate}
   \item Types of chart junk (arbitrary and schema junk);
   \item The individuation of attributes and the individuation of instances.
 \end{enumerate}

van Wijk's \cite{vanWijk:2006} model of visualization treats knowledge acquisition as the accumulation of knowledge elements over quantized intervals of time. The knowledge gained depends (a) on the existing knowledge of the user (a basic tenet of information theory) and (b) the perceptual and cognitive abilities of the user which will be affected by their prior experience (an acceptance of the post-structuralist view of context). The model also permits changing of the visualization specification (the layout $rules$ in our category) to facilitate further exploration of the data. Pinker \cite{Pinker:1990} can also be brought to bear here as his study of how people read graphs can be used to inform the design of layouts and data-to-representation mappings. A particular strength of van Wijk's work \cite{vanWijk:2006} is that it provides insight into why some types of visualization are successful and others not; it gives us a tool for understanding why a categorical description does not commute, that is, it helps us to understand how the $read$ transformation takes place to form Evocations in the mind of the user.

Hullman and Diakopoulos \cite{Hullman:2011} used the tools of rhetoric to offer techniques for understanding the selection of data (Schema and $measure$ment design), the design of representations (Layout), and the role of annotations and interactivity. They also deal explicitly with denotative and connotative meaning in understanding how visualizations are viewed. This aspect of their work finds its place on the intensional and extensional sides of our category.

\subsection{Summary}\label{sec:relevance}
The theoretical development in section \ref{sec:CTVis} gave rise to the visualization process category in Fig. \ref{fig:Process} and the summary of how different types of mapping affect a visualization through the $render$ morphism  in Table \ref{tab:morphSummary}. Fig. \ref{fig:Process} provides an overview of a well-formed visualization process. As far as a visualization designer is concerned, this category diagram will not tell them \emph{how} to create a visualization, rather it tells them all the major components that must be considered for correct knowledge creation and sense making to take place. Visualization design works principally at the intensional level while user interaction with an instance of the visualization process takes place at the extensional level.  Fig. \ref{fig:Process} shows the objects and morphisms but does not specify their types. Table \ref{tab:morphSummary} summarizes the main morphism types available and, as we showed in section \ref{sec:render}, these types materially affect the properties of the visualization. For example, when specifying a $render$ morphism to transform Data into a Representation, we saw that if the Representation was required to contain no ambiguity then it must be isomorphic (bijective for sets). Fig. \ref{fig:Process} also tells us that a complete visualization process needs to be considered from both the extensional (real world of instances) and intensional (definitional, schematic) levels. 

The category theory approach in this article takes a higher level view of visualization than the examples of previous work described above. These earlier theo\-retical and formal treatments, which focus either on taxonomic descriptions or on process to show how certain effects can be achieved using particular visualization techniques, can be seen to contribute to the development of one or more objects or morphisms within the visualization category in Fig. \ref{fig:Process}.  

\section{Conclusions}
\label{sec:conc}

This article is concerned with the foundational concepts necessary for developing from first principles a formal description of visualization. We have:
\begin{enumerate}
  \item Shown how the process of information visualization fits within the semiotic theory of Peirce and Saussure;
  \item Shown how the semiotic frameworks can be formally characterized using the language of category theory;
  \item Extended the Peircean semiotic triad into a mathematically closed category (see Fig.~\ref{fig:Process}) that incorporates the context of the visualization, including the 
  relationship between the triad and the original system under study, the relationship between the data, representation, schema, and layout, and the questions and human insight and knowledge that result from the visualization process;
  \item Used this formal framework to
  	\begin{enumerate} 
	   \item formally define visualization properties (literalness, redundancy, sensitivity, generalizability, chart junk) that were previously defined intuitively, if at all, and to 
	   \item expand on definitions provided in earlier work (see Table \ref{tab:morphSummary}) including differentiating between redundancy and arbitrary chart junk. These insights arise out of, and were discovered using, the category theoretic treatment of the visualization process; 
	\end{enumerate}
  \item Used the category to derive the property of 'intensionality' as applied to visualizations, and provided examples of how that property empowers users.
\end{enumerate}

Prior to this work many of the visualization concepts discussed above were understood intuitively or heuristically, but now we have verifiable formal theoretical descriptions that can be used to make principled judgements about the visualization process in general and about individual specific instances in particular. The category in Fig.~\ref{fig:Process} is a mathematically formal diagram that defines a well-formed visualization process; if all the morphisms and objects are well defined, the diagram will commute, and we can be confident that the knowledge gained is reliable and reflective of the system. The diagram can be used for reasoning and understanding. We can also envisage a tool that encapsulates the category's formalisms which, through some form filling or other exercise allows the user to specify details of the system, schema, and proposed layout, together with candidate questions. The tool could then validate the design and also show the consequences of specific design decisions. For example, a proposed layout would result in redundancy while another would exhibit non-ambiguity, sensitivity, etc.

We have identified the principal objects implicated in visualization,  the 14 morphisms (relationships)  that map between the structures in these objects, and four properties that each of those relationships may possess (monomorphism, epimorphism, isomorphism, and endo\-morphism). We have considered the implications to visualization by applying these properties to just one of those relationships (the $render$ morphism). 

In future work we will
consider the implications of these properties for the remaining morphisms and
apply higher-order category theory tools to compare different visualizations
  in terms of their ability to generate the same knowledge. This would be particularly appropriate when considering visualization across
  modalities allowing us to compare, say, a graphical representation with an auditory display.

\begin{IEEEbiography}[{\includegraphics[width=1in,height=1.25in,clip,keepaspectratio]{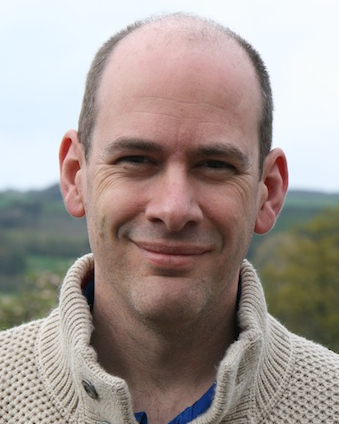}}]{Paul Vickers}
is a UK Chartered Engineer, holds a BSc degree in Computer Studies from Liverpool Polytechnic, and a PhD in Software Engineering \& HCI from Loughborough University. He is currently Reader in Computer Science at Northumbria University. His research is in the computing domain where it intersects with creative digital media with a particular emphasis on auditory display. He is a board member of the International Community for Auditory Display. 
\end{IEEEbiography}\vfill
\begin{IEEEbiography}[{\includegraphics[width=1in,height=1.25in,clip,keepaspectratio]{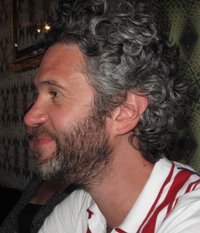}}]{Joe Faith}
has a BSc in Mathematics from Bristol University and a DPhil in Philosophy from Sussex University. He has combined academic research in visualization and bioinformatics with work in industry including as a founder of an eLearning company. The research in this article was undertaken while he was a Senior Lecturer at Northumbria University, and he is currently with Google Inc.
\end{IEEEbiography}\vfill
\begin{IEEEbiography}[{\includegraphics[width=1in,height=1.25in,clip,keepaspectratio]{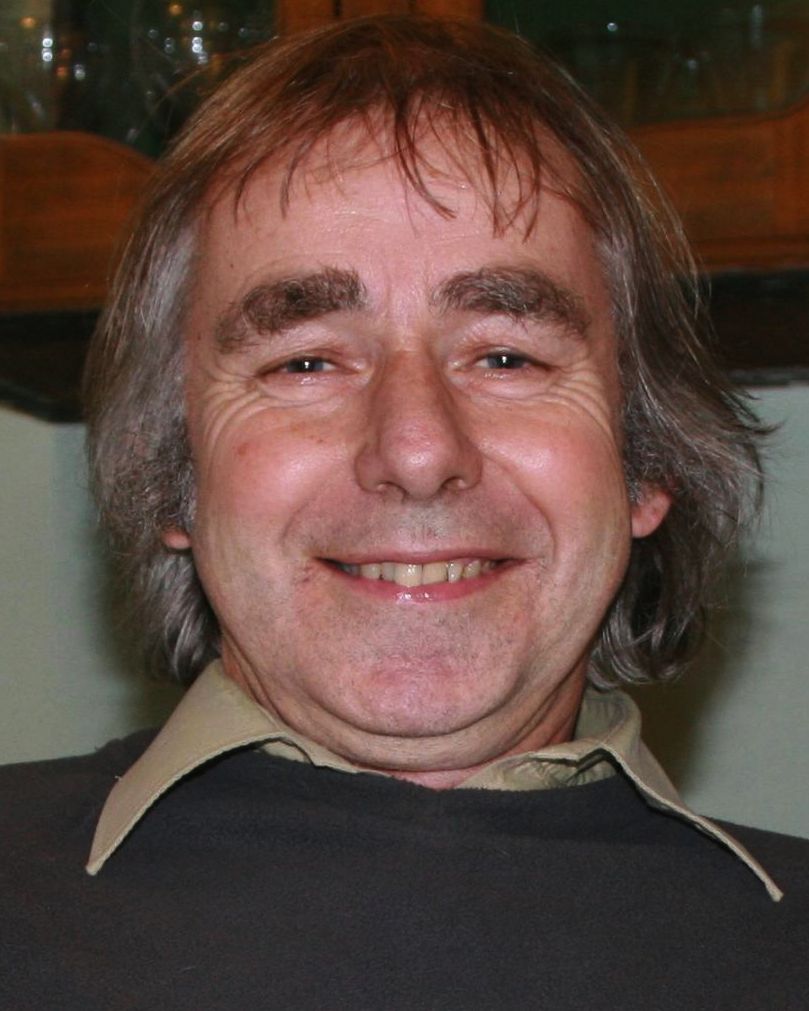}}]{Nick Rossiter}
has a BSc in Chemistry with Mathematics and a PhD in Physical Chemistry from Hull University. He is currently Visiting Research Fellow in Northumbria University's School of Computing, Engineering and Information Sciences, where he was formerly a Reader in Databases. His research interests include theoretical aspects of information systems and modelling in databases. 
\end{IEEEbiography}\vfill\vfill
%




\end{document}